\documentclass{article}
\usepackage{spconf}
\usepackage{comment}
\usepackage[T1]{fontenc}
\usepackage[polish]{babel}
\usepackage[utf8]{inputenc}
\usepackage{url}
\usepackage{hyperref}
\usepackage{amsmath,amssymb,amsfonts}
\usepackage{algorithmic}
\usepackage{graphicx}
\usepackage{textcomp}
\usepackage{xcolor}
\usepackage{tabularx}
\usepackage{booktabs}
\usepackage{multirow}
\usepackage{tikz}
\usepackage{ragged2e}
\usepackage{dblfloatfix}
\usepackage{flushend}
\usepackage{arydshln}
\usepackage{makecell}
\usepackage{lineno}
\usepackage{makecell}
\usepackage{cite}
\usepackage{placeins}
\usepackage{orcidlink}

\title{The Hidden Cost of Digits: Number Normalization and WER in ASR Systems}
\name{Stanisław Kacprzak\,\orcidlink{0000-0002-8717-9327} \qquad Mieszko Fraś\,\orcidlink{0000-0000-0000-0000}}

  \address{AGH University of Krakow, Institute of Electronics, Krakow, Poland \\}
\ninept
\begin{document}
\raggedbottom
\maketitle
\begin{abstract}
Modern automatic speech recognition (ASR) systems trained on extremely large datasets can produce transcripts with numbers written in Arabic numerals. This creates a need for fair comparison with models that output verbatim texts and proper processing of reference transcripts.
Popular approaches often reduce text normalization to lowercase and remove punctuation, with no additional normalization applied to languages other than English.
In this work, we analyze the impact of normalization of numerical expressions in the evaluation of ASR systems in various languages, using Polish as an example of a highly inflective language. 
We perform experiments on VoxPopuli and The Polish Parliamentary speech datasets and estimate word error rate (WER) differences for different text normalization approaches. We show that the difference due to the lack of number normalization in WER may be substantial - more than 2 percentage points, and often higher than the differences between systems in popular multilingual benchmarks. 
\end{abstract}
\begin{keywords}
automatic speech recognition, text normalization, number normalization, word error rate
\end{keywords}

\section{Introduction}
Advances in deep learning have led to large improvements in automatic speech recognition (ASR) in recent years.
The most common way to measure the improvement in ASR quality is the word error rate (WER), and in clean speech benchmarks \cite{open-asr-leaderboard} its value is getting closer and closer to zero. However, in \cite{szymanski2020we} the authors argue that the reported results are overoptimistic%
call for the creation of new benchmarks that are more aligned with the contemporary application domains of ASR systems. 
A critique of blindly relying on WER values was also presented in \cite{favre2013automatic}, where the authors suggest that the usefulness of the results obtained to a specific task should be measured, rather than strictly the quality of the transcript (which for many tasks is not crucial).
The need for more challenging benchmarks results in a lack of strict control over the data used for evaluation of the systems. 
Moreover, the increasing scale of training data for ASR systems has resulted in models that have learned to map between utterances and their transcribed forms, which can include punctuation marks, numbers, and abbreviations.
This makes a proper evaluation of the quality of system prediction a challenge.

Since the introduction of the Whisper model, its shared code \cite{whisper} is commonly used for text normalization (which is a transformation to some predefined standardized form), often ignoring that for languages other than English, only basic normalization is performed. The problem of normalization of texts in the evaluation of multilingual ASR systems was already pointed out in \cite{rouditchenko2023comparison} and \cite{manohar2024lost}, where the authors focus on artificially improved performance caused by normalization for Indic scripts. 
In \cite{molapo2014number} the authors focused on the problem of number normalization in multilingual societies, where numbers can sometimes be pronounced in different languages.  In \cite{sproat2016rnn} the authors describe the challenging task of text normalization with a deep learning approach, arguing the need to combine it with the knowledge-based approach (to prevent problematic errors). Better results but similar conclusions, that knowledge-base rules are extremely useful, plus an extensive description of the problem of text normalization, can be found in \cite{zhang2019neural}.
The NVIDIA NeMo toolkit \cite{NeMo_a_toolkit} contains some rule-based tools for text normalization for a few languages other than english%
, but even the multilingual evaluations of NVIDIA models in \cite{puvvada2024less, chen2024bestow, huang2025nest, canaryv2} and the Open ASR Leaderboard \cite{open-asr-leaderboard} are based on Whisper normalization.

In this work, we focus on text normalization with respect to numbers and show how much it may affect the WER estimation on a common benchmark. We see it as a complement to \cite{huber2024handling}, where the authors address the problem of correctly formatting numeric expressions in ASR transcripts.
Motivated by the observation that some systems output numeric expressions, while others produce verbatim transcripts, leading to inconsistencies in the WER evaluation.   Moreover, data sets can also contain transcripts in both forms, making it difficult to assess at a glance whether a particular WER result is over- or under estimated.
Although the problem is briefly mentioned in some previous works (such as \cite{whisper}, \cite{earnings}) and is probably well known to speech recognition engineers responsible for deploying ASR systems in production, it is still mostly neglected or overlooked (especially for languages other than English) in recent academic works and benchmarks, where fair and reliable comparisons of systems are sometimes overshadowed by the pursuit of beating the current state-of-the-art results. 
Furthermore, for many languages other than English, there is a lack of effective, proven, publicly available tools for number normalization.

The main contributions of our work consist of experiments that compare multiple normalization scenarios for various languages, reference transcript types, and multiple popular multilingual ASR systems, followed by an in-depth analysis of number normalization in English (for which publicly available tools allow both normalization and inverse normalization) and Polish language (in which due to its high inflection, it is a much more difficult task). Additionally, we propose our own rule-based words-to-digit normalizer\footnote{\label{our_implementation}\href{https://github.com/stachu86/polish-number-normalization}{Polish number normalizer repository}}.
Next, we present a possible WER improvement from the use of number normalization in different languages for popular ASR systems. Finally, we show that neglect of number normalization leads to differences of up to 2\% WER, and may cause misleading system comparison.

\section{Text normalization}
\label{sec:Text normalization}

\begin{table*}[t]
\centering
\caption{Examples of high WER obtained for ``perfect'' predictions caused by imperfect normalization algorithm from \cite{whisper}}
\label{tab:normalisation-errors}

\small

\resizebox{\textwidth}{!}{%
\begin{tabular}{llllr}
\textbf{Reference} & \textbf{Prediction} & \textbf{Normalized Reference} & \textbf{Normalized Prediction} & \textbf{WER} \\
\midrule
The meeting is at 10 am. & The meeting is at 10 a.m. & the meeting is at 10 am & the meeting is at 10 a m & 29\% \\
It's twelve twenty-five. & It's 12:25. & it is 1225 & it is 12 25 & 66\% \\
Width is 0.01 mm. &Width is one hundredth millimeter. & width is 0.01 & width is 100th of a millimeter & 133\% \\
One hundred, two, one, one. & 100, 2, 1, 1. & 10211 & 100 2 one one & 400\%\\
Thousand, million, two, thousand, two. & 1000, 1000000, 2, 1000, 2. & 3002 & 1000 1000000 2 1000 2 & 500\% \\
\end{tabular}%
}
\end{table*}

Text normalization is a process of converting different text representations of the same speech utterance into a more convenient standard form, and it is crucial for the evaluation of ASR systems. Performed basic normalization includes lowercase text, removal of punctuation, and for non-English texts handling of diacritics.
More specifically, the \emph{BasicTextNormalizer} used in \cite{whisper} performs: removal of any phrases between the matching brackets ([, ]) and parentheses ((, )), replacement of any markers, symbols, and punctuation characters with a space, transformation of the text to lowercase and replacement of any successive whitespace characters with a space. Moreover, it puts a space between every letter for languages that do not use spaces to separate words (e.g., Chinese). For the case of texts in English, \emph{EnglishTextNormalizer} performs normalization  with additional transformations (see Appendix C in \cite{whisper}) such as conversion of contracted forms, removal of commas between digits, removal of periods not followed by a number, conversion of British spellings into American spellings, and replacement of numeric expressions of numbers and currencies with a form using Arabic numbers (using \emph{EnglishNumberNormalizer}), but no additional transformations are applied to languages other than English.

The conversion of verbatim numerals is generally considered \emph{inverse text normalization}, but in this work we follow \cite{whisper} and consider it as a part of the normalization process. 
It is important to remember that this approach is far from perfect and that not so far-fetched edge cases can generate unexpected results. In Table \ref{tab:normalisation-errors} we provide some examples of, one would think, perfect predictions that result in high WER, due to imperfect and sometimes surprising normalization. 
In those examples, we see problems caused by a dot in "a.m", colon in time format, treating abbreviation of millimeter as a filler word and the problems of ambiguity and wrong assumptions when processing series of numbers (caused by removal of commas at the first step). 
Depending on the context in which the ASR system works and the frequency of those cases, the distortion of the evaluation results can vary.
However, the biggest issue is the lack of language-specific normalization when evaluating multilingual models.
 
\subsection{Polish Number Normalization}
Since Polish is a highly inflected language, building a rule-based normalizer is a difficult task because the correct form depends on the context. The detailed description of problems related to the normalization of Polish texts can be found in \cite{gralinski2006linguistic}. %
There are works exploring neural models to perform number normalization like \cite{poswiata2019numbers}, but to our knowledge, there are no freely available reliable models.

However, the inverse normalization is easier but requires a listing of all possible forms (or well crafted regular expressions). 
For this reason, we decided to follow the approach from \cite{whisper} and normalize verbatim transcriptions of numbers into expressions written with Arabic numbers. Our \emph{PolishNumberNormalizer}\footnotemark[1] implementation is based on OpenAI's official code.
We added support for Polish conventions (use of commas instead of dots), inflections of ordinal numbers, handling of numbers with decimal fractions, dates and time, and perform number normalizations before removal of punctuation (thus reducing possible ambiguity). 
Our solution is not able to normalize all possible ordinal numeral inflections or fractions (like proper fractions), but we believe that the most common cases are properly processed and problems of Whisper's \emph{EnglishNumberNormalizer} described in Table~\ref{tab:normalisation-errors} are fixed. 
The code contains unit tests for typical use cases, and the correctness of the \emph{PolishNumberNormalizer} was validated on the Polish corpus of parliamentary speech \cite{sejmsenat} (see Subsection \ref{normalization_impact}).

\subsection{Number Normalization influence on evaluation}
We recall the definition of WER as
\begin{equation}
    WER[\%] = 
    \frac{S + I + D}
    {N}
    \times 100\%\,,
\end{equation}
where, S is the number of substitutions, D is the number of deletions, I is the number of insertions and N is the total number of words in the reference transcription. 
It is important to note that normalizing verbatim numerals into a single number, results in reduction of the total number of words N, which increases the weight of errors on total WER, and similarly normalization into verbatim form will decrease the weight of errors. 
The difference in approaches can be observed in the results presented in \cite{whisper}, where the authors showed differences in WER obtained using different normalizers, one that performs number normalization and the other that performs inverse number normalization. 
This difference may vary depending on the language, which can be observed in Table~\ref{tab:1984}, where we show an example of how many words are needed to represent the number 1984 verbatim depending on the language.

\section{Datasets, ASR systems and experimental setup}
\label{sec:datasets}
In this section, we describe the datasets and ASR systems used in our work, as well as the performed experiments.

\subsection{Datasets}
For English, French, Spanish, German and Polish languages, we used VoxPopuli \cite{voxpopuli} dataset. In addition, we used the Polish corpus of parliamentary speech \cite{sejmsenat}, where we suspect a large number of numerals (dates, amounts of money, statistics, and numbers of legal acts).

\begin{table}[h]
\caption{Example of verbatim form of number 1984 in different languages. Words are counted after normalization whith \emph{BasicTextNormalizer}\cite{whisper}}
\label{tab:1984}
\begin{tabular}{cp{0.7\columnwidth}c}
\textbf{Lng.} & \textbf{Verbatim number} & \makecell{\textbf{Number}\\\textbf{of words}} \\
\hline
DE & neunzehnhundertvierundachtzig & 1 \\
EN & nineteen eighty-four & 3 \\
PL & tysiąc dziewięćset osemdziesiąt cztery & 4 \\
ES & mil novecientos ochenta y cuatro & 5 \\
FR & mille neuf cent quatre-vingt-quatre & 6 \\
\end{tabular}
\end{table}

VoxPopuli contains two types of transcript \emph{raw\_text} (original audio segment text) and \emph{normalized\_text} (normalized audio segment transcription). The \emph{raw\_text} contains numbers mostly written with Arabic digits, while \emph{normalized\_text} contains only verbatim numbers; therefore, for clarity, in the remainder of this paper we will refer to them as \emph{digit text} and \emph{verbatim text}, respectively. In our analysis, we used only test splits with approximately two thousand recordings for selected languages: English (1842), Polish (1831), German (1968), French (1742), Spanish (1528).

The Polish Parliamentary dataset contains the original transcription with verbatim numbers (\emph{verbatim text}). Since the test set is relatively small (130 recordings), we manually created the \emph{digit text}. 
During the analysis of the ASR outputs, we identified two recordings for which the transcripts did not correspond to the audio content and excluded them from the experiments.

\subsection{Automatic Speech Recognition Systems}
\label{subsec:ASR}
In our experiments, four current state-of-the-art ASR systems were used: Whisper \cite{whisper} large-v3, Canary 1B \cite{puvvada2024less}, Canary 1B-v2 \cite{canaryv2}, and OWSM v4 medium \cite{owsmv4}.
Importantly for our analysis, the Whisper and OWSM systems output primarily \emph{digit text}, Canary-1B-v2 outputs mostly \emph{verbatim text}, and Canary-1B outputs only \emph{verbatim  text}. Those differences come from the way the systems were trained. %
Furthermore, for the Whisper model, we can influence the model to produce verbatim transcription following two approaches: suppression of digit tokens or using an initial prompt. During our initial experiments, we noticed that the second approach yields better results, so in this work, for some of the experiments, we use the following prompt: \emph{"Today is May twenty-third, twenty twenty-five. The meeting started at three thirty PM. About seventy-five percent of the invited guests attended. The report covered one hundred twenty-three cases. The results were in line with expectations."}, for the English system and translated version for experiments with other languages. In all of the experiments, we explicitly choose the recognition language and use beam search decoding with beam size set to five.

\subsection{Performed experiments}
In order to investigate the actual importance of number normalization, we performed the following experiments.
First, we analyze the actual and maximum possible WER between different types of transcript (\emph{digit text} and \emph{verbatim text}) and how much it can be reduced using number normalization for English and Polish. 
Then, we calculate the WER for the predictions for each system and analyze the impact of number normalization on both the output of the system and the references.
Finally, for all languages and systems, we calculate the WER gain obtained by normalizing both reference transcripts and ASR output.
Note that in the aforementioned experiments, all transcripts and ASR outputs were processed with a \emph{BasicTextNormalizer} from \cite{whisper} (described in Section \ref{sec:Text normalization}), to remove capital letters, punctuation, and symbols other than letters and digits.
However, to avoid ambiguities associated with punctuation (as in the examples in Table~\ref{tab:normalisation-errors}), we first apply the number normalizer and then \emph{the BasicTextNormalizer} is applied to the resulting text.
For number normalization in English, we use \emph{EnglishNumberNormalizer} from \cite{whisper}, for Polish datasets, we use our \emph{PolishNumberNormalizer} described %
in Section \ref{sec:Text normalization}, while for German, French, and Spanish we use NeMo \emph{InverseNormalizer} \cite{NeMo_a_toolkit, nemo_int}.
In either case, we denote the number normalization operation as N($\cdot$). 
We explicitly refrain from using \emph{EnglishTextNormalizer}, to only focus on impact of the number normalization; consequently, the results may differ from those presented in other works, e.g \cite{puvvada2024less} or in Open ASR Leaderboard\cite{open-asr-leaderboard}.
\begin{table}[h!]
    \caption{
    WER [\%] between digit (D) and verbatim (V) transcripts.
    }
    \label{table:transcripts}
    \centering
    \resizebox{\columnwidth}{!}{
        
    \begin{tabular}{c|l|c|cccc}
Dataset & Reference & Digits [\%] & V & D & N(V) & N(D) \\
\hline
        
        \hline
        \multirow{4}{*}{\rotatebox[origin=c]{90}{\shortstack{English \\ VoxPopuli}}}
        & V & 0.00 & -- & 0.98 & 1.21 & 1.32 \\
        & D & 0.33 & 0.98 & -- & 0.70 & 0.40 \\
        & N(V) & 0.68 & 1.22 & 0.70 & -- & \textbf{0.31} \\
        & N(D) & 0.62 & 1.33 & 0.40 & \textbf{0.31} & -- \\

        \hline
        \multirow{4}{*}{\rotatebox[origin=c]{90}{\shortstack{Polish \\ VoxPopuli}}}
        & V & 0.00 & -- & 1.27 & 1.55 & 1.79 \\
        & D &  0.51 & 1.28 & -- & 0.88 & 0.60 \\
        & N(V) & 0.95 & 1.56 & 0.88 & -- & \textbf{0.30} \\
        & N(D) & 0.93 & 1.80 & 0.60 & \textbf{0.30} & -- \\

        \hline
        \multirow{4}{*}{\rotatebox[origin=c]{90}{\shortstack{Polish \\ Parliamentary}}}
        & V & 0.00 & -- & 2.17 & 2.84 & 2.86 \\
        & D & 1.76 & 2.19 & -- & 0.72 & 0.70 \\
        & N(V) & 1.21 & 2.87 & 0.72 & -- & \textbf{0.00} \\
        & N(D) & 1.78 & 2.89 & 0.70 & \textbf{0.00} & -- \\
    \end{tabular}
    }
\end{table}

\begin{table*}[h!]
    \centering
    \caption{
    WER in \% obtained on various datasets with and without number normalization
        }
    \label{table:results}
\setlength{\tabcolsep}{5pt}
\renewcommand{\arraystretch}{1.05}
\begin{tabular}{@{}c|l|cccccccccc@{}}
    \multirow{2}{*}{Dataset} &
    \multirow{2}{*}{Reference} &
    \multicolumn{2}{c}{Whisper} &
    \multicolumn{2}{c}{Whisper-IP} &
    \multicolumn{2}{c}{OWSM} &
    \multicolumn{2}{c}{Canary-v2} &
    \multicolumn{2}{c}{Canary-v1} \\
    
    & &
     & N($\cdot$)&
     & N($\cdot$)&
     & N($\cdot$)&
     & N($\cdot$)&
     & N($\cdot$)\\
    \hline
        \multirow{4}{*}{\rotatebox[origin=c]{90}{\shortstack{English \\ VoxPopuli}}}
        & V  & 8.46 & 8.77 & 8.38 & 9.03 & 8.02 & 8.34 & 7.50 & 7.98 & 5.95 & 6.73  \\
        & N(V) & 8.24 & \bf{7.87} & 8.99 & \bf{8.36} & 7.77 & \bf{7.40} & 7.64 & \bf{7.13} & 6.46 & \bf{5.64}\\
        \cdashline{2-12}
        & D  & 7.65 & 8.01 & 8.45 & 8.54 & 7.22 & 7.58 & 7.09 & 7.31 & 5.95 & 5.98\\
        & N(D) & 8.01 & \bf{7.64} & 8.79 & \bf{8.19} & 7.57 & \bf{7.21} & 7.45 & \bf{6.95} & 6.29 & \bf{5.59}\\

        \hline
        \multirow{4}{*}{\rotatebox[origin=c]{90}{\shortstack{Polish \\ VoxPopuli}}}
        & V  & 7.85 & 8.28 & 7.52 & 8.51 & 13.86 & 14.27 & 7.95 & 8.78 \\
        & N(V) & 7.45 & \bf{6.97} & 8.52 & \bf{7.23} & 13.47 & \bf{13.02} & 8.57 & \bf{7.51} \\
        \cdashline{2-10}
        & D  & 7.16 & 7.64 & 8.09 & 7.83 & 13.24 & 13.68 & 8.20 & 8.20  \\
        & N(D) & 7.57 & \bf{7.08} & 8.62 & \bf{7.28} & 13.61 & \bf{13.16} & 8.72 & \bf{7.64}\\

        \cline{1-10}
        \multirow{4}{*}{\rotatebox[origin=c]{90}{\shortstack{Polish \\ Parliamentary}}}
        & V  & 15.34 & 15.76 & 12.73 & 15.12 & 29.87 & 30.33 & 17.49 & 17.97 \\
        & N(V) & 14.03 & \bf{13.32} & 15.02 & \bf{12.52} & 28.64 & \bf{28.11} & 16.19 & \bf{15.45} \\
        \cdashline{2-10}
        & D  & 13.69 & 13.93 & 14.41 & 13.18 & 28.46 & 28.67 & 15.86 & 16.10 \\
        & N(D) & 14.01 & \bf{13.30} & 15.02 & \bf{12.50} & 28.64 & \bf{28.09} & 16.19 & \bf{15.43} \\
    \end{tabular}
\end{table*}

\section{Experimental results and discussion}
\label{sec:experiments}
In this section, we present and discuss the results of our experiments and analyses.
\subsection{Impact of number normalization on analyzed datasets}
\label{normalization_impact}
First, we analyze the impact of numerical expressions on two types of transcripts. 
The percentage of words containing digits and the WER obtained between the \emph{digit text} and the \emph{verbatim text} transcripts and these transcripts after number normalization are presented in Table \ref{table:transcripts}.
In the case of the English VoxPopuli dataset, applying number normalization significantly increases the percentage of words with digits, even for the \emph{digit text}.
Note that switching from verbatim to digit numerals decreases total number of words, which means that other words have a greater impact on WER.
Moreover, normalization introduces WER of 0.40\% and 1.22\% for \emph{digit text} and \emph{verbatim text}, respectively.
Most importantly, if both transcripts were normalized, WER decreased to 0.31\%, while in the most pessimistic scenario, in which only one transcript is normalized, WER reaches 1.33\%.
Remaining differences (not offset by number normalization) mostly consist of word splitting (e.g. "well being" $\rightarrow$ "wellbeing", "money laundering" $\rightarrow$ "moneylaundering") and a few individual other errors, like missing words, in original \emph{digit text} transcript.

The same analysis was performed on the VoxPopuli Polish dataset, which contains a significantly higher percentage of words with digits. 
As expected, the WER differences are also higher compared to the English dataset. 
More importantly, number normalization allows to decrease the WER between \emph{digit text} and \emph{verbatim text} transcripts from initial 1.28\% to 0.30\%.
Observed WER of 0.60\% between the \emph{digit text} and the \emph{Norm(digit text)} show that even the 'raw' transcript with numbers should be further normalized for the fair system comparison.
Additionally, we conducted this analysis on The Polish Parliamentary dataset, as we expected it to contain a high number of numerical values in various contexts. Its relatively small size also allowed for a more in-depth and insightful analysis and allowed us to verify all predictions and manually perform the normalization of the numbers for \emph{digit text}.
As expected, the percentage of words containing digits is much higher (almost double) than in the Polish VoxPopuli dataset.
Moreover, the WER between \emph{digit text} and \emph{verbatim text} is greater than 2\%, but is reduced to zero, which demonstrates the high effectiveness of the proposed \emph{PolishNumberNormalizer}.

Summarizing the above results, in the pessimistic scenario, we estimate the maximum WER associated with the lack of normalization of the numbers for the English VoxPopuli dataset to be above 1\%, while reaching 1.5\% and almost 3\% for Polish VoxPopuli and Polish Parliamentary datasets, respectively.
\subsection{Impact of number normalization on ASR evaluation}
In the next experiment, we examine the impact of number normalization (or lack thereof) on the resulting WER using state-of-the-art multilingual ASR systems. 
The results obtained with all systems, with and without number normalization, are presented in Table \ref{table:results}.
Whisper \cite{whisper} large-v3 is described as \emph{Whisper}, while \emph{Whisper-IP} denotes the same system, but using the initial prompt to output verbal numbers, as described in Section~\ref{subsec:ASR}. The OWSM v4 medium \cite{owsmv4}, Canary 1B \cite{puvvada2024less}, and Canary 1B-v2 \cite{canaryv2} systems are denoted \emph{OWSM}, \emph{Canary-v1}, and \emph{Canary-v2}, respectively.

First, one can observe that in the case of both English and Polish VoxPopuli datasets, Whisper-IP performs worse than unprompted Whisper in normalized scenarios by about 0.5\% WER, which we attribute to the fact that prompting generally slightly degraded the results.
However, despite being a poorer performing system, without number normalization it achieved better results in verbatim text than standard Whisper (8.38\% WER vs 8.46\% WER in English and 7.52\% vs 7.85\% for Polish), which clearly highlights the need for the use of number normalization.
Most importantly, for each system, transcript and dataset, the lowest WER can be obtained if both the transcript and the system prediction are number normalized (bold for each combination of transcript and system).
Interestingly, in the case of the Polish Parliamentary dataset, while overall results are much worse for all systems, \emph{Whisper-IP} seems to produce more reliable predictions than the basic version. One of the reasons that explains this behavior is that using prompt caused the Whisper model to produce verbatim numbers in places where Roman numerals were used in \emph{Whisper} predictions, e.g. "XXVIII posiedzenie Sejmu" (28th sitting of the Sejm). Nevertheless, the best results for all combinations of transcripts and systems are still obtained when both transcriptions and system output are number-normalized. 

\begin{figure}[t!]
    \centering
    \includegraphics[width=1\columnwidth]{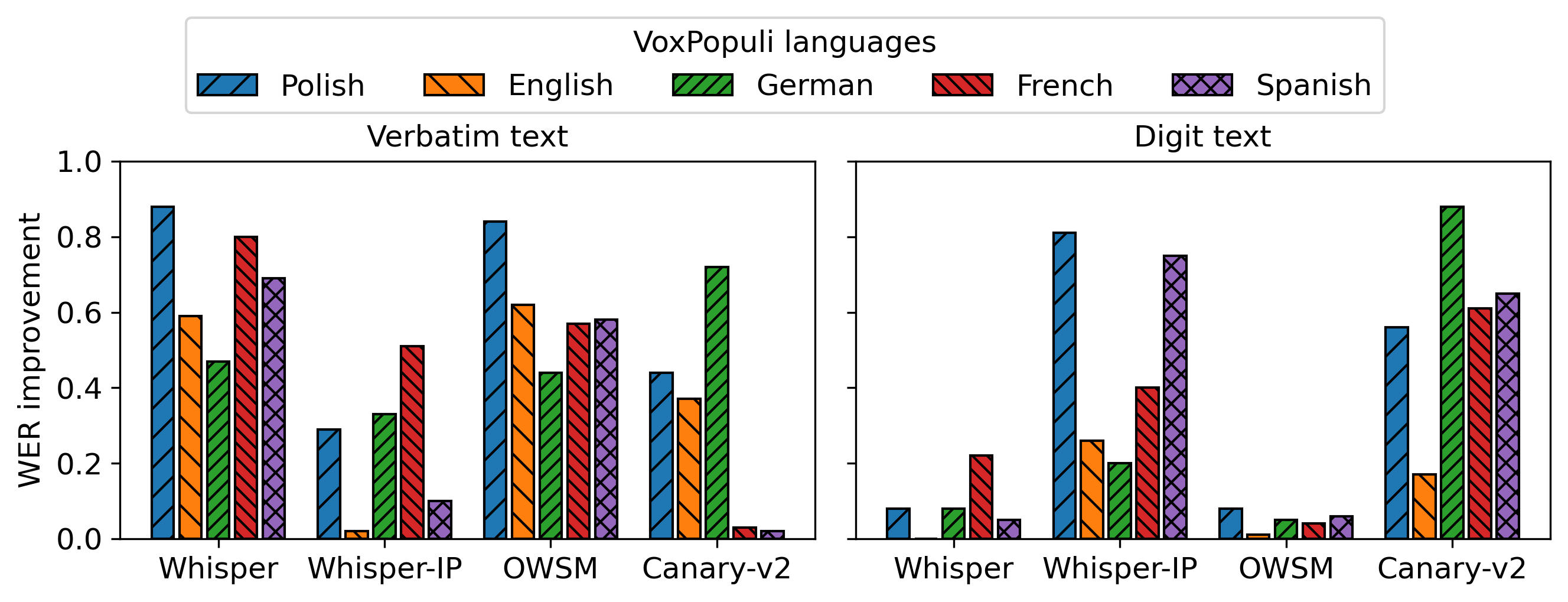}
    \caption{Absolute WER reduction (percentage points) in WER when ASR outputs and reference transcripts are both number normalized for various VoxPopuli languages}
    \label{fig:wykres}
\end{figure}

The summary of the gains obtained by the number normalization of both the reference transcripts and the ASR outputs, for various languages from the VoxPopuli datasets, is presented in Figure \ref{fig:wykres}.
\subsection{Observations and discussions}
We observe gain in terms of WER due to number normalization, for all languages, especially for combination of verbatim reference and digit-returning systems, and the opposite combination. However, in some cases, e.g. German and \emph{Canary-v2} system, notable gains are visible for both transcripts.
These results confirm that it is very difficult to estimate the impact of number normalization on a given system in a given language.
According to the results of our experiments, for some systems, more than 10\% of the relative WER may be attributed to the lack of proper normalization of numbers in certain languages.
We believe that introducing a proper normalization pipeline could provide a more fair comparison for multilingual ASR systems, especially since the differences between the best-performing models are often very small, in the order of tenths of a percent WER \cite{open-asr-leaderboard}.%

\section{Conclusions}
\label{sec:conclusions}
In this work, we analyze the impact of number normalization on the WER calculation, an aspect that is commonly omitted in the case of the evaluation of ASR systems, especially in languages other than English. 
We point out some of the problems of commonly used English normalizer, and provide an implementation of normalizer for the Polish language\footnotemark[1]. 
The experimental results presented show that, depending on the language and dataset, artificially induced discrepancies between systems that output numerals as digits versus verbatim lead to differences in WER of up to 1 percentage point on VoxPopuli dataset and up to 2 percentage points on Polish Parliamentary dataset.
Most importantly, we have shown that the lack of number normalization can produce a misleading system comparison. This suggests that special care must be taken when comparing models that predict text in different standards.
\clearpage
\subsection*{Acknowledgments}
This research was supported by the Excellence Initiative – Research University Programme
for the AGH University of Krakow and, in part, by the National Science Centre, Poland under Grant 2021/42/E/ST7/00452. For the purpose of Open Access, the author has applied a CC-BY public copyright licence to any Author Accepted Manuscript (AAM) version arising from this submission.
\subsection*{Generative AI Use Disclosure}
Generative AI tools were used in two limited roles. ChatGPT (OpenAI) was used to provide editorial and critical feedback on the clarity, presentation, and argumentation of the manuscript, while GitHub Copilot was used to assist with code review and minor code improvements. All suggestions were evaluated by the authors and any generated code was reviewed, modified where necessary, and tested. The tools were not used to generate experimental results, validate scientific claims, or make scientific decisions. The authors take full responsibility for the manuscript, the reported results, and the correctness of the software.
\bibliographystyle{IEEEbib}
\bibliography{mybib}

\end{document}